\documentclass[aps,prc,twocolumn,groupedaddress,showpacs,showkeys]{revtex4}
\usepackage[dvips]{graphicx}
\usepackage{bm}
\usepackage{epsfig}
\usepackage{amsfonts}
\usepackage{amssymb,amscd}
\usepackage{subfigure}
\usepackage{latexsym}
\usepackage{amsmath}
\usepackage{slashed}
\usepackage{xcolor}

\newcommand{\be}{\begin{equation}}
\newcommand{\ee}{\end{equation}}
\newcommand{\bea}{\begin{eqnarray}}
\newcommand{\eea}{\end{eqnarray}}
\newcommand{\bem}{\begin{multline}}
\newcommand{\eem}{\end{multline}}
\newcommand{\beg}{\begin{gather}}
\newcommand{\eeg}{\end{gather}}

\newcommand{\gt}{>}

\newcommand{\ben}{\begin{eqnarray*}}
\newcommand{\een}{\end{eqnarray*}}

\graphicspath{{Figs/}}

\begin{document}

\title{From \bm{$D_{s}^{\pm}$} production asymmetry at the LHC 
to prompt \bm{$\nu_{\tau}$} at IceCube
}

\author{Victor P.~Goncalves}
\email{barros@ufpel.edu.br}
\affiliation{Instituto de F\'{\i}sica e Matem\'atica,  Universidade
Federal de Pelotas (UFPel), \\
Caixa Postal 354, CEP 96010-900, Pelotas, RS, Brazil}

\author{Rafa{\l} Maciu{\l}a}
\email{rafal.maciula@ifj.edu.pl} \affiliation{Institute of Nuclear
Physics, Polish Academy of Sciences, Radzikowskiego 152, PL-31-342 Krak{\'o}w, Poland}

\author{Antoni Szczurek\footnote{also at University of Rzesz\'ow, PL-35-959 Rzesz\'ow, Poland}}
\email{antoni.szczurek@ifj.edu.pl} \affiliation{Institute of Nuclear
Physics, Polish Academy of Sciences, Radzikowskiego 152, PL-31-342 Krak{\'o}w, Poland}


\begin{abstract}
The description of the heavy meson production at large energies and forward rapidities at the LHC is fundamental to derive realistic predictions of the prompt atmospheric neutrino flux at the IceCube Observatory. In particular, the prompt tau neutrino flux is  determined by the decay of $D_s$ mesons produced in cosmic ray - air interactions at high energies and large values of the Feynman - $x_F$ variable. Recent data from the LHCb Collaboration indicate a production asymmetry for  $D_s^+$ and $D_s^-$ mesons, which cannot be explained in terms of the standard modelling of the hadronization process. In this paper we demonstrate that this asymmetry can be described assuming an asymmetric strange sea ($s(x) \ne \bar s(x)$)  in the proton wave function and taking into account of the charm and strange fragmentation into $D_s$ mesons. Moreover, we show that the strange quark fragmentation contribution is dominant at large - $x_F$ ($\ge 0.3$). The prompt $\nu_{\tau}$ flux is calculated  and the enhancement  associated to the strange quark fragmentation contribution, disregarded in previous calculations, is estimated.
\end{abstract}


\maketitle

{\it{Introduction.}}
The experimental results obtained in recent years by the LHC, the Pierre
Auger and  IceCube Neutrino Observatories have challenged and improved
our understanding of Particle Physics. The discovery of the Higgs boson
\cite{higgs} completed the Standard Model (SM), which is now a complete
and self-consistent theory. On the other hand,  the detection of
astrophysical neutrinos by the  IceCube Neutrino Observatory sets the
begining of the neutrino astronomy \cite{IceCube_Science}. In addition,
the data from the Pierre Auger Observatory provide an unique oportunity to test Particle Physics at energies well beyond current accelerators \cite{auger}. Such results motivated, in particular,  the development of new and/or more precise approaches to describe the perturbative and nonperturbative regimes of the Quantum Chromodynamics (QCD). One example is the recent improvement in the description of the heavy meson production in hadronic collisions at the LHC, directly influenced by the need to constrain the magnitude of the prompt neutrino flux, which is crucial for a precise determination of the cosmic neutrino flux at the IceCube 
\cite{calc_recentes,anna,GMPS2018,prdandre}. In what follows we will explore this direct connection between the LHC and IceCube results and provide more precise predictions for the $D_s$ production at the LHC and the prompt tau neutrino flux at the IceCube.

The atmospheric neutrinos are produced in  cosmic-ray interactions with nuclei
in Earth's atmosphere \cite{review_neutrinos}. At low neutrino energies ($E_{\nu}\lesssim 10^5$ GeV), these neutrinos arise from the decay of light mesons (pions and kaons), and the associated flux is denoted as the {\it conventional} atmospheric neutrino flux 
\cite{Honda:2006qj}. On the other hand, in the 
energy range 10$^{5}$ GeV $< E_{\nu} <$ 10$^{7}$ GeV, it is expected
that the {\it prompt} atmospheric neutrino flux associated with the
decay of hadrons containing heavy flavours become important \cite{ingelman}.
In the particular case of the tau neutrino $\nu_{\tau}$ flux, it is
dominated at low energies by the conventional atmospheric flux, via
$\nu_{\mu} \rightarrow \nu_{\tau}$ oscillations. On the other hand, for
$E_{\nu} >$ 10$^{4}$ GeV, 
this contribution becomes negligible and the prompt $\nu_{\tau}$ flux 
is determined by the decay of $D_s$ mesons, which have a leptonic decay 
channel $D_s \rightarrow \tau \nu_{\tau}$ with a branching ratio of a few percent, with the subsequent $\tau$ decay that also contributes to the flux \cite{PR}. A precise determination of the prompt $\nu_{\tau}$ flux is fundamental to identify the tau neutrinos of cosmic origin, which is considered another important signature of the cosmic ray origin of the highest neutrino flux. 
As demonstrated in Ref. \cite{GMPS2018}, the prompt neutrino flux is determined by the heavy meson production at high energies and very forward rapidities. 
Therefore, the description of the $D_s$ production in the kinematical 
region probed by the LHCb Collaboration is a requisite to obtain a precise prediction of the prompt $\nu_{\tau}$ flux. 

During the last years, the LHCb Collaboration released a large set of
data associated with the $D$ and $B$ meson production. The data for the
transverse momentum and rapidity distributions are, in general, quite
well described by theoretical approaches. On the other hand, the description of the experimental data for the charge production asymmetries 
\cite{LHCb:2012fb,Aaij:2018afd} still is a challenge for the great majority of the theoretical approaches. In general, these production asymmetry are interpreted as arising during the nonperturbative process of hadronization as implemented e.g. in the PYTHIA Monte Carlo, which is based on the Lund string fragmentation model. However, this approach fails to describe the recent LHCb data for the $D_s^{\pm}$ production asymmetry, which have found evidence of a nonzero asymmetry. In particular, in contrast with PYTHIA that predicts a positive value, the experimental data indicate that the asymmetry is negative. Very recently, two of us have proposed in Ref.~\cite{Maciula:2017wov} an alternative approach to describe the asymmetry present in the $D^+$ and $D^-$ production \cite{LHCb:2012fb}. The basic idea is that subleading contributions for the parton fragmentation are nonnegligible at the LHC energies and that the asymmetry comes
from the inherent asymmetry of the $u$ and $d$ valence distributions in the incident protons. In this paper we extend the approach for the $D_s^{\pm}$ production and  demonstrate that the LHCb data can be described if we assume that the strange quark sea in the proton is asymmetric, with $s \neq \bar{s}$. Such asymmetry is predicted e.g. by the Meson Cloud Model (see. e.g. Refs. 
\cite{Holtmann,Feng:2012gu,CGNN2013,WJMSTW2016,iran}) and is not
excluded by the recent data and by QCD global analysis. In reality, the
strange sea in the proton is only poorly known, with its behavior being
determined in a great 
extent by the neutrino - induced DIS data on charm production obtained by the CCFR/NuTeV and NOMAD experiments 
\cite{CCFR,NuTeV}. Recent analysis of the LHC data for the $W^{\pm}$
production improved our understanding of the strange distribution,
especially at small - $x$, but the existence or not of an asymmetric
strange sea is still an open question \cite{amanda}. As a consequence,
assuming that our approach for the subleading parton fragmentation is
correct, the results for the $D_s^{\pm}$ 
asymmetry can be considered as a first signature that $s \neq \bar{s}$ in the proton.
Finally, the results presented in Ref.~\cite{Maciula:2017wov} and in what follows indicate that the subleading contributions to the parton fragmentation are nonnegligible and have a large impact on the Feynman - $x_F$ distributions of the heavy mesons produced in hadronic collisions. As discussed e.g. in Refs. \cite{anna,GMPS2018}, this distribution determines the prompt neutrino flux. Therefore, it is expected that the prompt $\nu_{\tau}$ flux will be enhanced by these subleading contributions. One of the goals of this paper is to estimate this enhancement and provide realistic predictions for the tau neutrino flux  that are based on a formalism that is in agreement with the recent LHCb data.

{\it $D_s$ production at the LHC.} At high energies the charm quarks are produced  dominantly by gluon - gluon interactions via the $gg \rightarrow c \bar{c}$ subprocess and are believed to hadronize to $D_s$ - mesons mainly through the $c \rightarrow D_s$ fragmentation process. Therefore, at leading order, we expect an identical amount of $D_s^+$ and ${D_s^-}$ mesons, which implies  that the charge asymmetry  defined by
\begin{equation}
A_P (D_s^+) = \frac{\sigma(D_s^+) - \sigma(D_s^-)}{\sigma(D_s^+) + \sigma(D_s^-)}
\label{asymmetry}
\end{equation}
will be zero at this approximation. Consequently, the asymmetries are expected to be generated by   subleading partonic subprocesses, initial state asymmetries and/or   a distinct description of the hadronization process.
Here we extend the approach proposed in
Ref.~\cite{Maciula:2017wov}, which explains the $D^+/D^-$ asymmetry in
terms of the unfavored fragmentation functions, which are responsible
for light quark/antiquark fragmentation to $D$ mesons, for $D_s$ production. The fact that $u \neq d$ in the incident protons, naturally leads to the $D^+/D^-$ asymmetry when the subleading contribution for the fragmentation is taken into account.
In contrast, in the case of $D_s$ production, the inclusion of the 
 subleading fragmentation, associated to the $s \to D_s^-$ and $\bar s
 \to D_s^+$ transformations, implies $A_P (D_s^+) = 0$ if $s =
 \bar{s}$. Therefore, our interpretation of the LHCb data is that the
 $D_s^{\pm}$ asymmetry arises due to an asymmetry in the strange quark
 sea. As explained before, such a behavior is predicted by some
 theoretical models and is not excluded by the recent QCD global
 analysis of the experimental data. In particular, the CTEQ
 Collaboration has performed a dedicated study of the strange parton
 distribution of the proton in Ref. \cite{Lai:2007dq} and obtained that
 the experimental data is quite well described also assuming that $s \neq \bar{s}$. 
 In the present paper we shall include the dominant $g + g \rightarrow  c + \bar{c}$  subprocess as well as the  $s/\bar s + g \to s/\bar s + g$ and $g + s/\bar s \to g + s/\bar s$ terms with the strange partons from Ref.~\cite{Lai:2007dq}. Moreover, we will include the subleading $s/\bar s \to D_s^{\pm}$ fragmentation, which is described in terms of the probability of transition of a strange quark into a $D_s$ meson ($P_{s \to D_s}$). Of course, the leading fragmentation is associated with the
$c \to D_s^+$ and $\bar c \to D_s^-$ transitions with an associated
transition branching of about 5-9 \% \cite{Lisovyi:2015uqa}.
As in  Ref.~\cite{Maciula:2017wov} the gluon - gluon channel will be calculated within $k_T$-factorization approach, with the cross section being expressed in terms of the  unintegrated gluon distribution function (UGDF) and  off-shell matrix element
for the $g g \to c \bar c$ subprocess. In the present paper we use 
the Kimber-Martin-Ryskin (KMR) prescription for UGDF \cite{Watt:2003mx}
which was shown to allow good description of charm production at 
the LHC \cite{Maciula:2013wg}.
In contrast, the $s(\bar s) g \to s(\bar s) g$ and $g s(\bar s) \to g s(\bar s)$ processes are calculated within a leading-order collinear approach, with the regulation of small transverse momenta region being done as in Ref.~\cite{Maciula:2017wov}.
Moreover, the $c \to D_s$ fragmentation is described using the Peterson model (with  $\varepsilon$ = 0.05), while the  $s \to D_s^-$ and $\bar s \to D_s^+$ fragmentation functions are parametrized as
in Ref.~\cite{Maciula:2017wov} using the reversed-Peterson and triangular 
fragmentation functions. The only free parameter in our approach 
 is $P_{s \to D_s}$. In principle, it is expected to be larger than the value obtained in Ref.~
\cite{Maciula:2017wov} for the $u/d \rightarrow D$ transition, due to the larger  mass of the strange quark. However, as this quantity is associated with a nonperturbative process, it is not possible to calculate its value from first principles. In what follows, we will  constrain   $P_{s \to D_s}$ using the LHCb data for the charge production asymmetry $A_P (D_s^+)$.

\begin{widetext}

\begin{figure}[t]
\begin{tabular}{cc}
\includegraphics[width=0.42\textwidth]{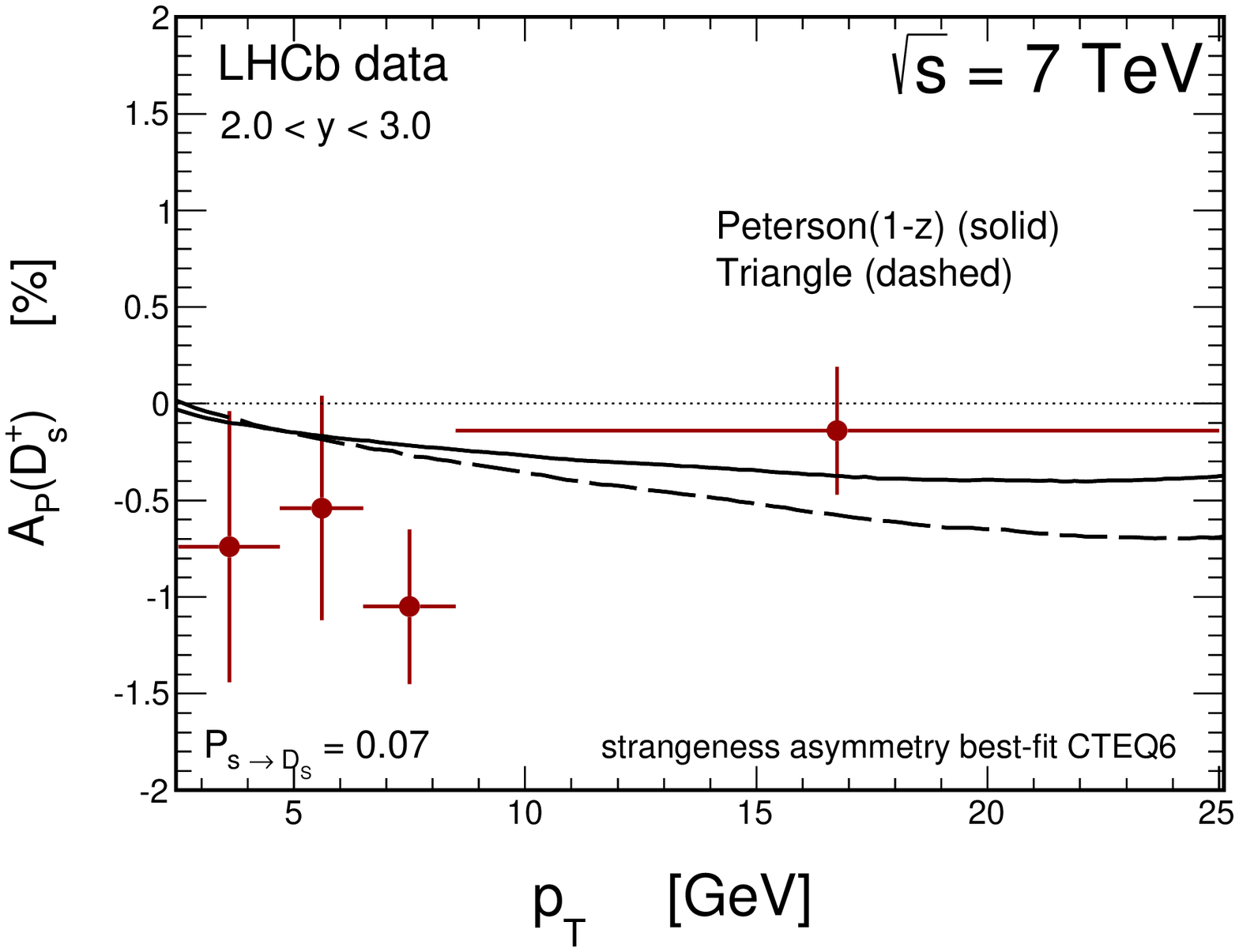} &
{\includegraphics[width=0.42\textwidth]{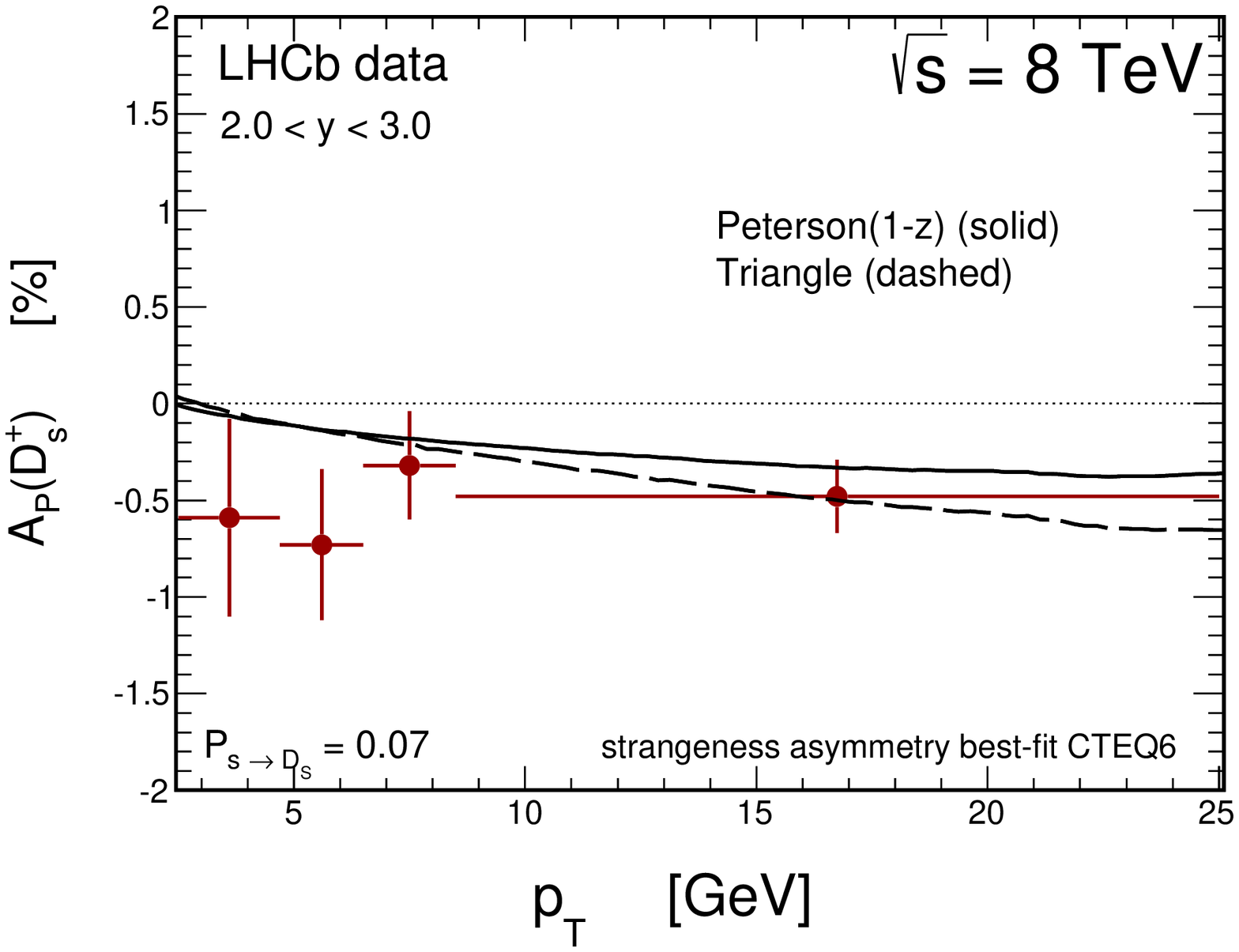}} \\
{\includegraphics[width=0.42\textwidth]{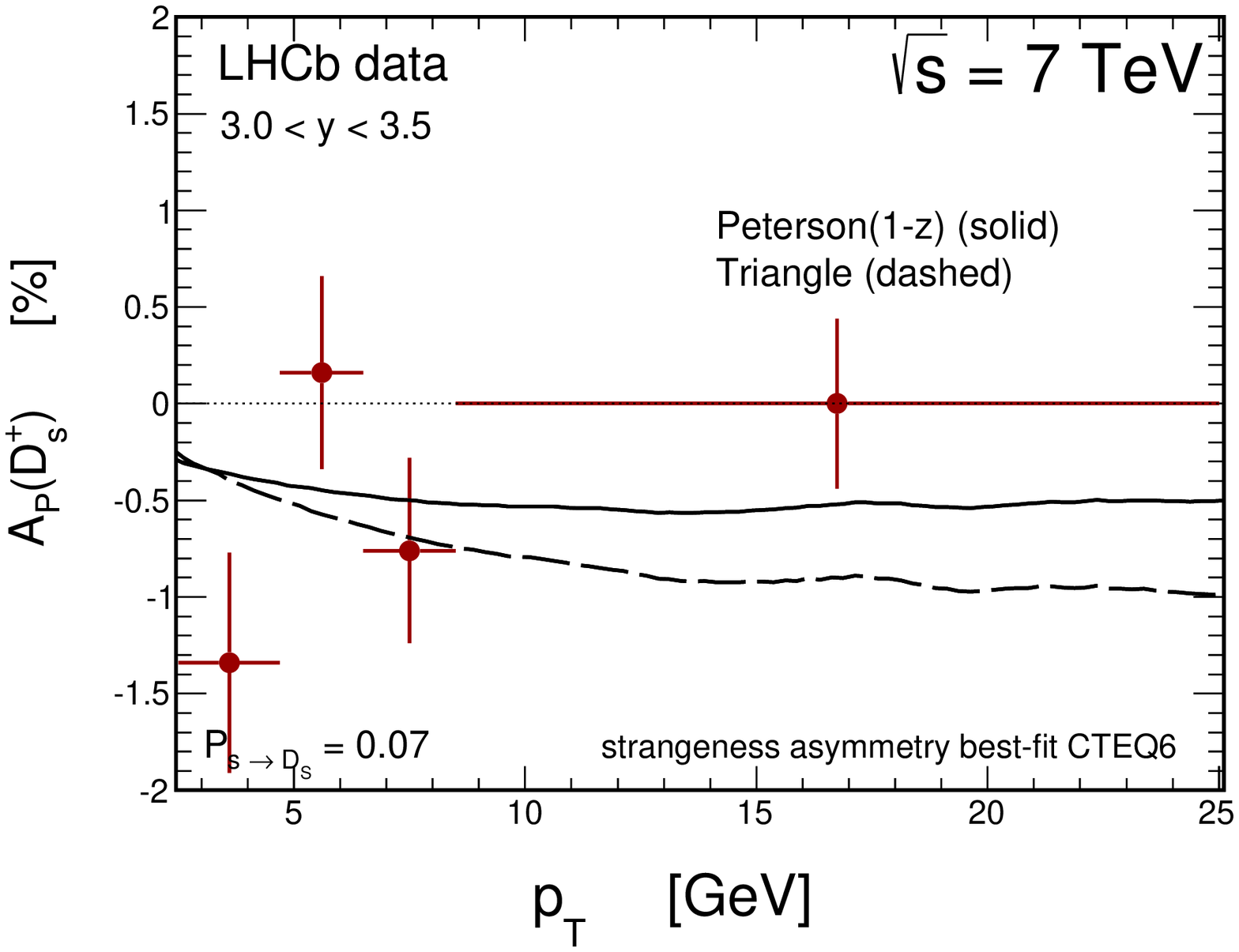}} &
{\includegraphics[width=0.42\textwidth]{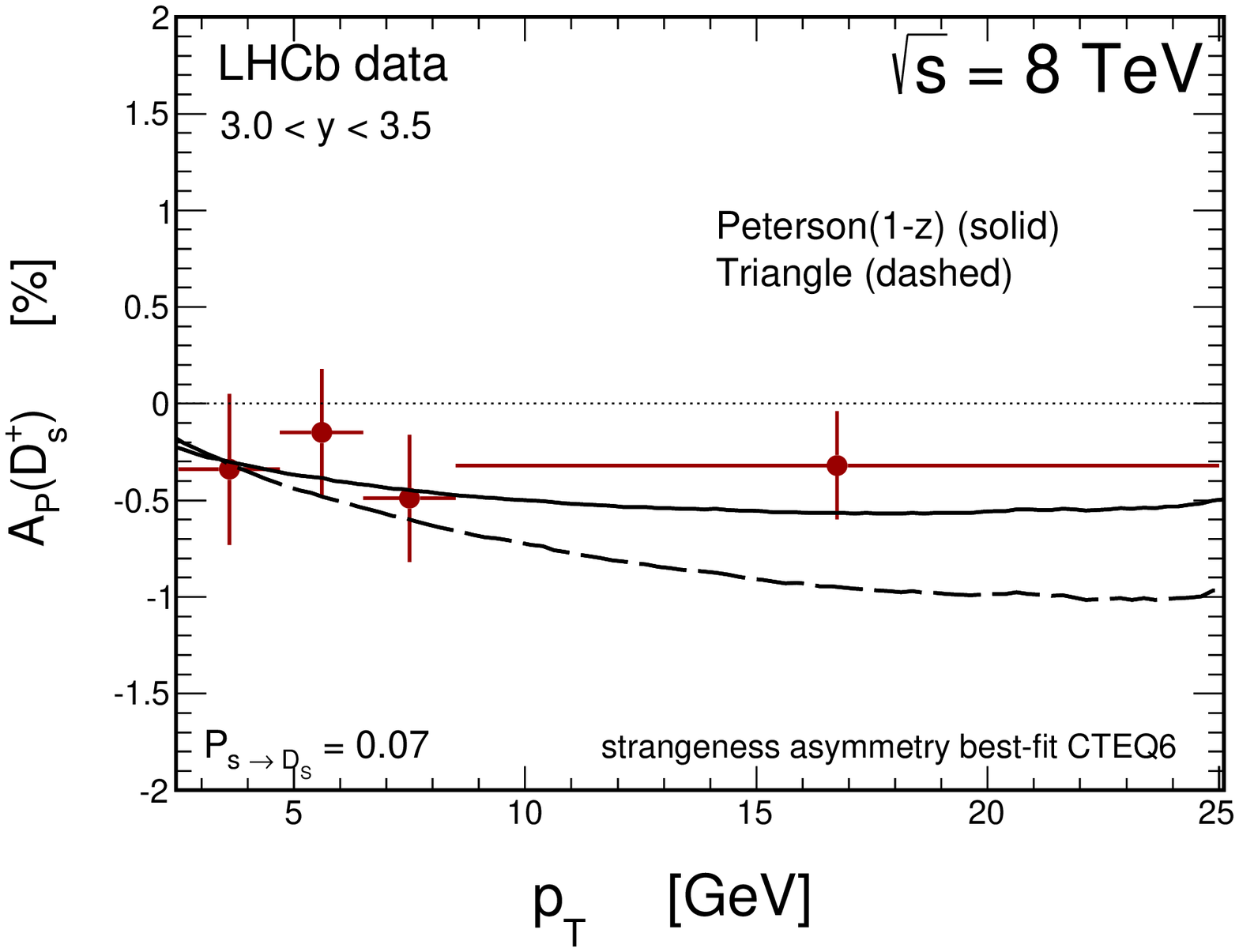}}\\
{\includegraphics[width=0.42\textwidth]{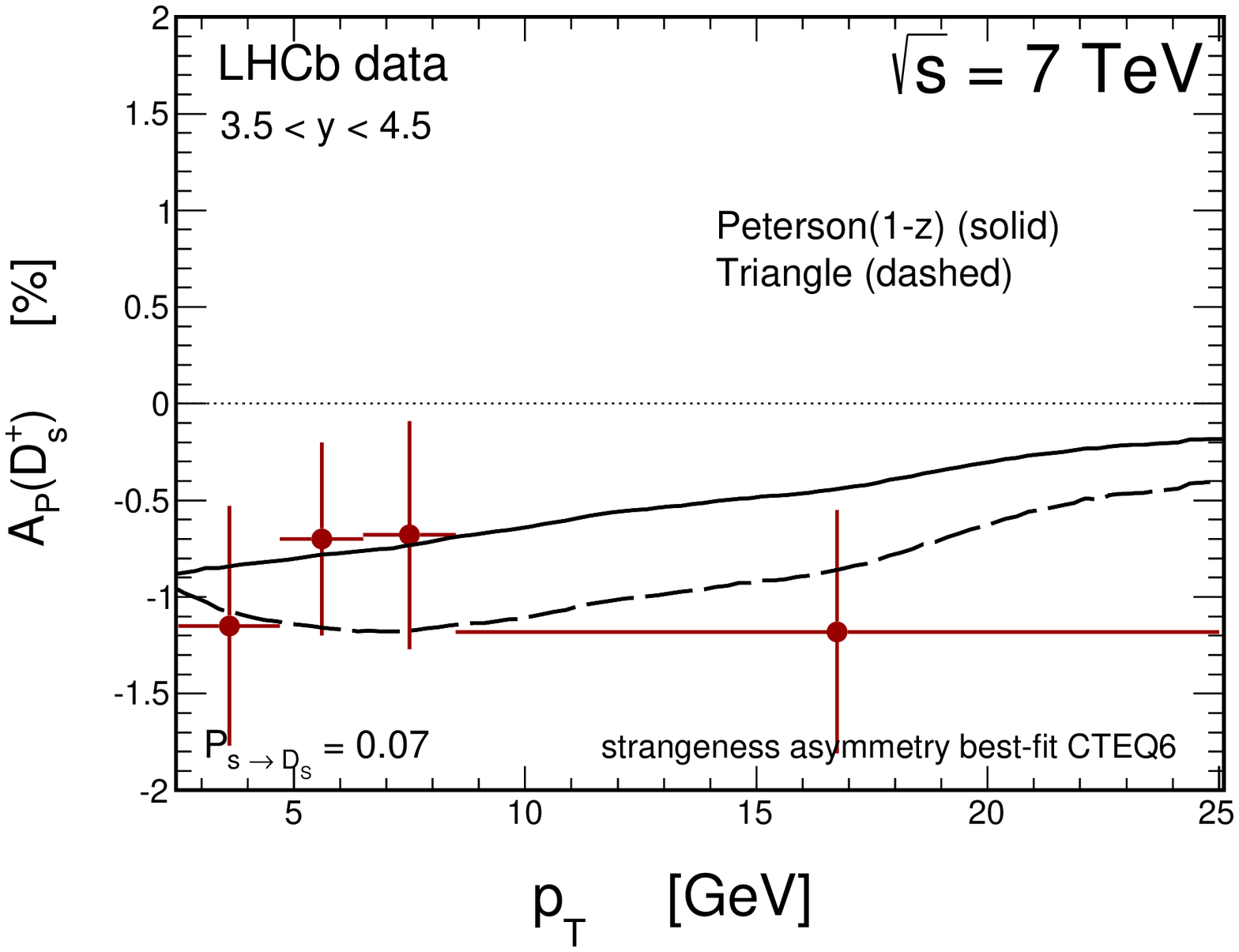}} &
{\includegraphics[width=0.42\textwidth]{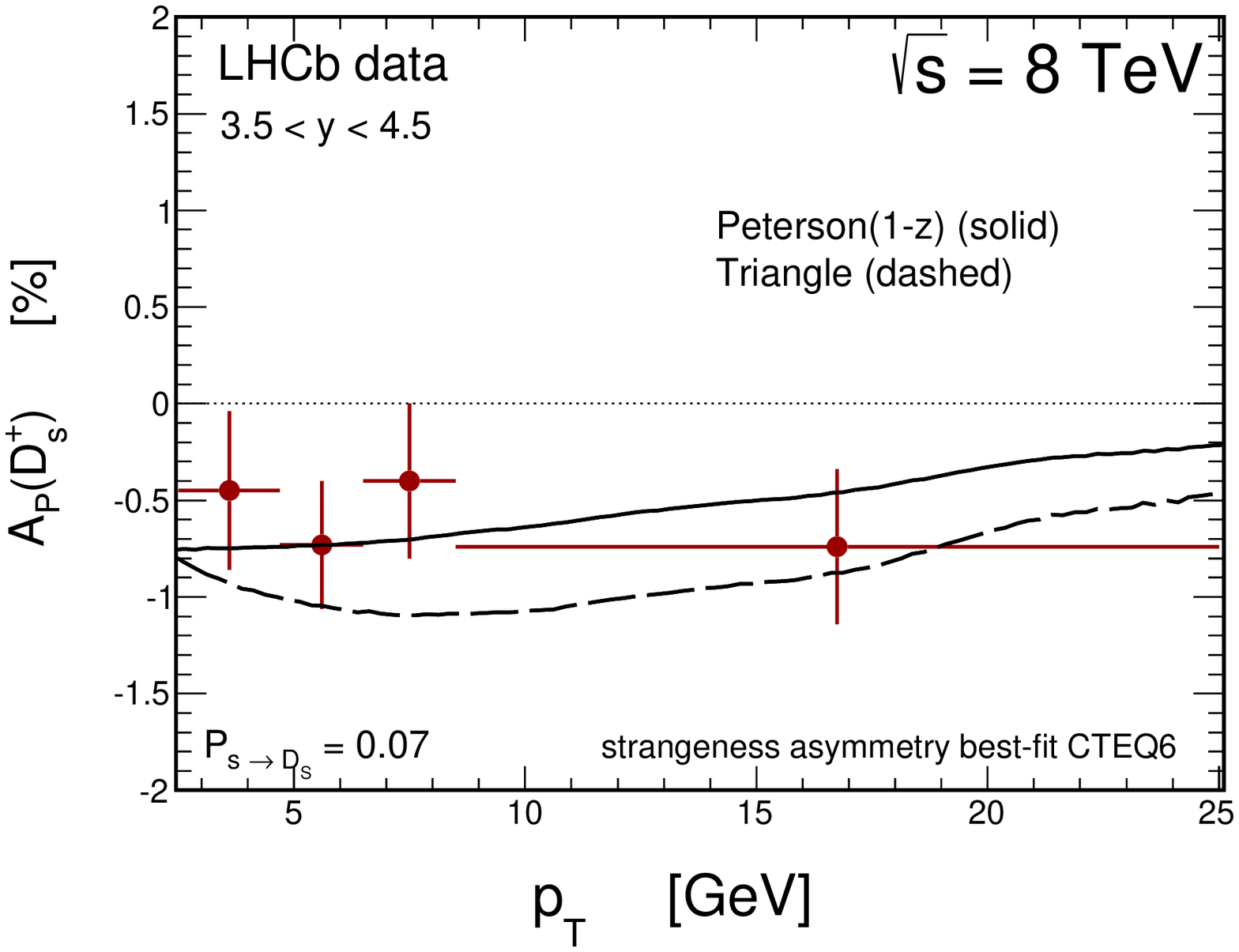}}
\end{tabular}
 \caption{$D_s^+ - D_s^-$ asymmetry obtained with our approach
for $\sqrt{s}$ = 7 TeV (left panel) and $\sqrt{s}$ = 8 TeV (right
panel) together with the LHCb experimental data \cite{Aaij:2018afd}.
\small 
 }
 \label{fig:Ds_asymmetry}
\end{figure}

\end{widetext}

\begin{widetext}

\begin{figure}[t]
\begin{tabular}{cc}
\includegraphics[width=0.42\textwidth]{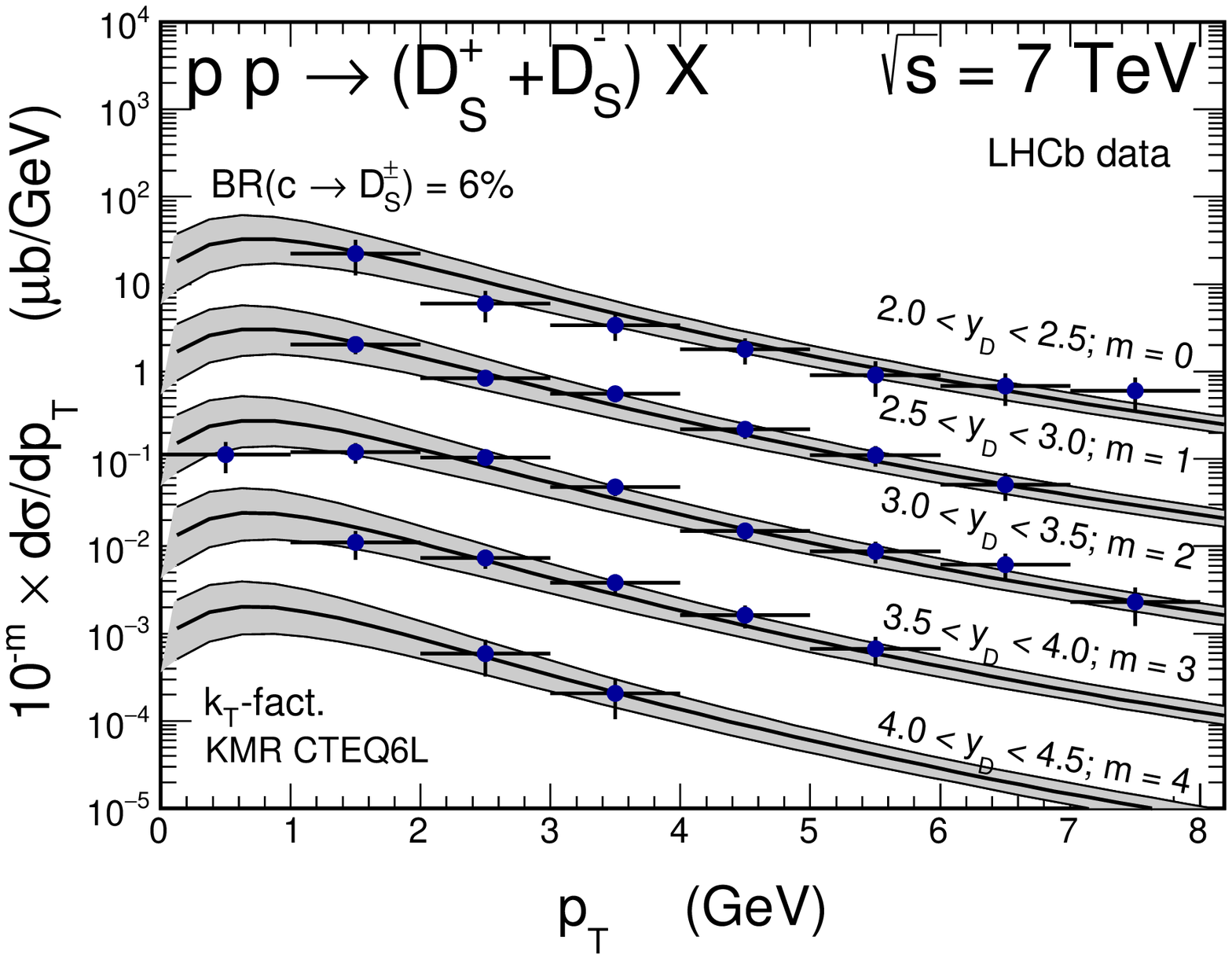} &
{\includegraphics[width=0.42\textwidth]{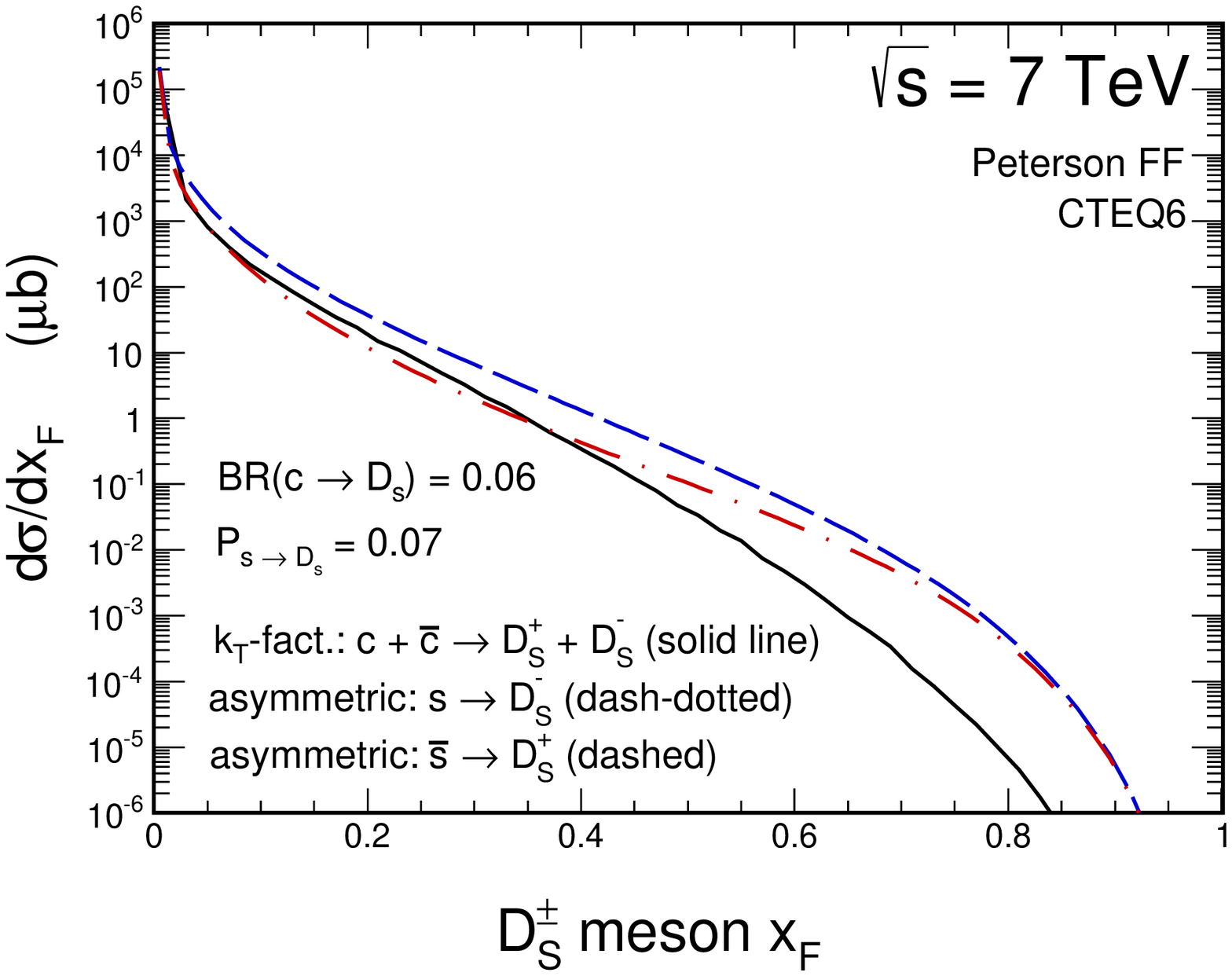}} 
\end{tabular}
 \caption{\small Left: Transverse momentum distributions of $D_s^+ + D_s^-$ for
different ranges of rapidity. The LHCb data \cite{Aaij:2013mga} are shown for comparison.
 Right: Feynman - $x_F$ distribution for the $D_s^{\pm}$ production.}
 \label{fig:Ds_meson}
\end{figure}

\end{widetext}

In Fig.\ref{fig:Ds_asymmetry} we present our results for the
$D_s^+ - D_s^-$ asymmetry for $\sqrt{s}$ = 7 TeV (left panels)
and $\sqrt{s}$ = 8 TeV (right panels) using the reversed-Peterson and triangular 
fragmentation functions for the $s \rightarrow D_s$ transition. Rather
reasonable agreement with the LHCb data is obtained assuming $P_{s \to
  D_s}$ = 7 $\cdot$ 10$^{-2}$. In other words, a  small value for the
unfavoured fragmentation function is sufficient to describe the LHCb
data.  The data statistics  is still too low to perform a more detailed fit and/or to discriminate between the two models for the subleading fragmentation. However, the results indicate that the asymmetry in strangeness in the proton wave function, as described in the CTEQ parametrization,  is able to generate the correct sign for  $A_P (D_s^+)$, in contrast to PYTHIA \cite{Aaij:2018afd}, as well as the enhancement of the asymmetry at larger rapidities. In Fig. \ref{fig:Ds_meson} (left panel) we present the resulting predictions for the  transverse momentum distributions of $D_s^+ + D_s^-$ for the different ranges of the meson rapidity ($y_D$) probed by the LHCb Collaboration \cite{Aaij:2013mga}.
A quite well agreement with the LHCb data  is achieved
without free parameters. We have verified that the contribution of the  subleading fragmentation for the $p_T$ - spectra is small ($\le 5 \%$) in the kinematical range probed by the LHCb Collaboration. In contrast, the behavior of the rapidity and Feynman - $x_F$ distributions are significantly modified at large values of $y_D$ and $x_F$. 
In Fig. \ref{fig:Ds_meson} (right panel) we demonstrate that the asymmetry in the strange sea imply different behaviors for the $x_F$ - distributions of the $D_s^+$ and $D_s^-$ mesons at intermediate $x_F$. More important, 
while at small - $x_F$ the conventional contribution dominates, at large - $x_F$ the situation is reversed. One has that 
the contribution associated to the 
$\bar s g \to \bar s g$ channel becomes dominant for $x_F \ge 0.05$.  In particular, for $x_F \gtrsim 0.3$, the channels initiated by strange quarks, usually disregarded in the analysis of  the $D_s$ production, are dominant. Such values of $x_F$ correspond to rapidities larger than those probed by the LHCb detector. However, as demonstrated in Ref. \cite{GMPS2018}, this is exactly the kinematical region that determines the behavior of the prompt neutrino flux. Consequently, the presence of the subleading contributions for the $D_s$ production is expected to have direct impact on the predictions of the prompt tau neutrino flux.

{\it Prompt $\nu_{\tau}$ flux at the IceCube.} One of the current goals
of the IceCube Observatory is the measurement of  tau-neutrinos
\cite{Aartsen:2015dlt}, which are considered an independent probe of the
cosmic neutrinos. Such an expectation is strongly motivated by the fact that 
for cosmic neutrinos the decay of charged pions generated in
astrophysical sources implies a ratio $\nu_e : \nu_{\mu} : \nu_{\tau}$ =
1 : 1 : 1 at the Earth, while for atmospheric neutrinos this ratio is expected to be typically
$\nu_e : \nu_{\mu} : \nu_{\tau}$ = 1 : 1: 0.1 \cite{anna,PR}. As a consequence, the background associated to atmospheric tau neutrinos is usually predicted to be strongly reduced in comparison to the other flavours, with the measurement of a tau neutrino being considered a direct probe of cosmic neutrinos. However, previous analysis have disregarded the subleading contributions discussed in this paper.
It is the aim of the present study to make a realistic estimate of the
prompt $\nu_{\tau}+\overline{\nu}_{\tau}$ flux when using the current
information from the LHC.

In order to estimate the prompt $\nu_{\tau}+\overline{\nu}_{\tau}$ flux, we will closely follow the  procedure described  in detail in Refs. \cite{GMPS2018,prdandre}. We will calculate the prompt tau neutrino flux  
using the semi-analytical $Z$-moment approach \cite{ingelman}, where  a set of coupled cascade  equations for the nucleons, heavy mesons and leptons (and their antiparticles) 
fluxes is solved, with the equations being expressed in terms of the nucleon-to-hadron
($Z_{NH}$), nucleon-to-nucleon ($Z_{NN}$), hadron-to-hadron  ($Z_{HH}$) 
and hadron-to-neutrino ($Z_{H\nu}$) $Z$-moments. 
These moments are inputs in the calculation of the prompt tau neutrino flux
associated with the production of a  $D_s$ meson and its decay into a
$\nu_{\tau}$ in the low- and high-energy regimes. We will focus on 
vertical fluxes  and will 
assume that the cosmic ray flux $\phi_N$  can be described by the H3a spectrum proposed in Ref. \cite{gaisser}, with the incident flux being
represented by protons. As in Ref. \cite{PR} we will include in our
calculations the contribution of neutrinos produced in the direct $D_s
\rightarrow \nu_{\tau}$ decay as well as those generated in the chain
decay $D_s \rightarrow {\tau} \rightarrow \nu_{\tau}$. The contribution
for the prompt $\nu_{\tau}$ flux associated to the decay of
mesons heavier than t$D_s$ is negligible \cite{anna} and will not be included in our analysis.

\begin{widetext}

\begin{figure}[t]
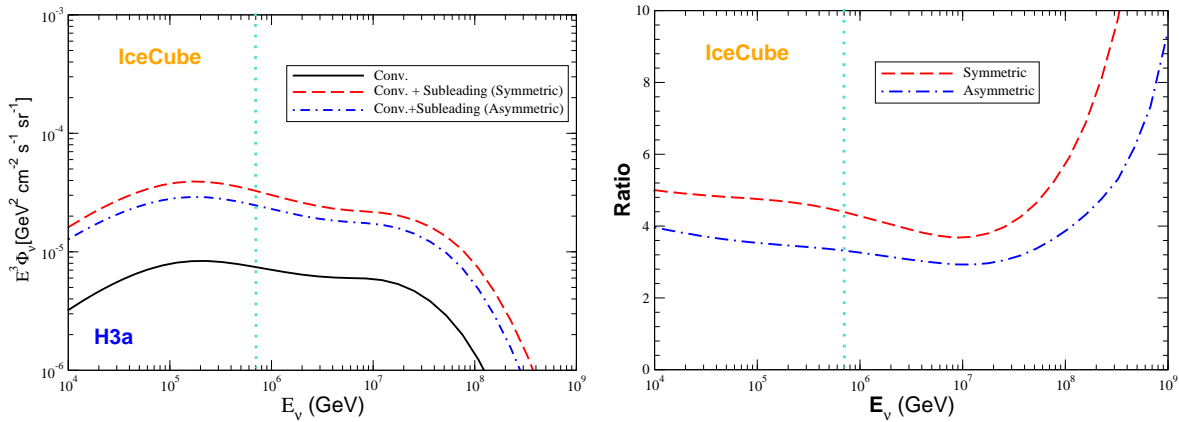

\begin{tabular}{ccc}
\includegraphics[width=0.43\textwidth]{tauneutrino.eps} & \, &
{\includegraphics[width=0.42\textwidth]{ratiotauneutrino.eps}} 
\end{tabular}
 \caption{\small Left: The prompt atmospheric tau neutrino ($\nu_{\tau} + \overline{\nu}_{\tau}$) 
flux as a function of the neutrino energy.  Right: Ratio between the full calculation (conventional + subleading contributions) and the conventional one.}
 \label{fig:spectra}
\end{figure}

\end{widetext}

In Fig.\ref{fig:spectra} (left) we show the flux
of the prompt $\nu_{\tau} + \overline{\nu}_{\tau}$ flux scaled 
by $E_{\nu}^3$.
In addition to the conventional component, associated to heavy quark production by a gluon-gluon fusion and represented
by the solid black line, we show the results which includes in addition the
subleading fragmentation assuming a symmetric (dashed red line) or an
asymmetric (dot - dashed blue line) strange sea in the proton wave
function. The subleading mechanism leads to a significant enhancement of
the high-energy prompt $\tau$-neutrino  flux, which can be quantified
calculating the ratio between the full prediction (conventional +
subleading) and the conventional one. The energy dependence of this
ratio is presented in Fig.\ref{fig:spectra} (right), which demonstrate
that the inclusion of the subleading contributions implies an
enhancement of a factor $\gtrsim 3.25$ in the prompt $\nu_{\tau}$ flux
in the kinematical range probed by the IceCube Observatory. In
particular, for a  asymmetric strange sea, needed to describe the LHCb
data, we predict that the enhancement will be of a factor of about 3.5 at $E_{\nu} = 10^5$ GeV.
The enhancement strongly increases for $E_{\nu} \gt 10^8$ GeV due to the faster decreasing of the gluon distribution in comparison to the strange one at large values of Bjorken $x$ variable.  


{\it Conclusions.} In the present paper we propose, for the first time, the description of the production asymmetry
for $D_s^+$ and $D_s^-$ mesons in terms of an asymmetry in the strange
sea of the proton associated to the inclusion of the subleading
fragmentation mechanisms $s \to D_s^-$ or $s \to D_s^+$. We have used
asymmetric $s - \bar s$ distributions derived in a global analysis of
different experimental data and that can be explained  within meson cloud
picture of the nucleon.  We have demonstrate that a small value for the
strange fragmentation function into $D_s$ mesons is sufficient to
describe the LHCb data. Such a new contribution, disregarded in previous
studies, becomes dominant at large values of $x_F$, that is the
kinematical range that determines the prompt
atmospheric $\nu_{\tau}$ flux at the IceCube Observatory. We have estimated the impact of this contribution and demonstrated that it implies an enhancement by a factor larger than 3 in the kinematical range probed by the IceCube Observatory. Our results indicate that a future experimental analysis of prompt tau neutrinos at the IceCube can be useful to probe the underlying mechanism of $D_s$ production at high energies and forward rapidities at the LHC.

We have found recently that the production of $\tau$-neutrinos was
discussed very recently in the context of intrinsic charm in the nucleon
\cite{BR2018} and the beam dump fixed target experiment SHiP at CERN
\cite{SHiP}. No reference to the LHCb asymmetry was done there.

{\it Acknowledgments.} 
This study was partially
supported by the Polish National Science Center grant
DEC-2014/15/B/ST2/02528, by the Center for Innovation and
Transfer of Natural Sciences and Engineering Knowledge in
Rzesz{\'o}w and  by the Brazilian funding agencies CNPq,  FAPERGS and  INCT-FNA (process number 464898/2014-5).


\end{document}